**Association for Information Systems**
**AIS Electronic Library (AISeL)**

PACIS 2017 Proceedings

Pacific Asia Conference on Information Systems (PACIS)

Summer 7-19-2017

# Addressing Knowledge Leakage Risk caused by the use of mobile devices in Australian Organizations

Carlos Andres Agudelo Serna
*The University of Melbourne*, cagudelo@student.unimelb.edu.au

Rachelle Bosua
*Department of Information Systems, The University of Melbourne, Victoria, Melbourne*, rachelle.bosua@unimelb.edu.au

Sean Maynard
*The University of Melbourne*, seanbm@unimelb.edu.au

Atif Ahmad
*The University of Melbourne*, atif@unimelb.edu.au

# Addressing Knowledge Leakage Risk caused by the use of mobile devices in Australian Organizations

*Completed Research Paper*

**Carlos Andres Agudelo-Serna**
School of Computing and Information Systems.
The University of Melbourne.
cagudelo@student.unimelb.edu.au

**Rachelle Bosua**
School of Computing and Information Systems.
The University of Melbourne.
Rachelle.bosua@unimelb.edu.au

**Sean B. Maynard**
School of Computing and Information Systems.
The University of Melbourne.
seanbm@unimelb.edu.au

**Atif Ahmad**
School of Computing and Information Systems.
The University of Melbourne.
atif@unimelb.edu.au

## Abstract

*Information and knowledge leakage has become a significant security risk to Australian organizations. Each security incident in Australian business cost an average US$2.8 million. Furthermore, Australian organisations spend the second most worldwide (US$1.2 million each on average) on investigating and assessing information breaches. The leakage of sensitive organizational information occurs through different avenues, such as social media, cloud computing and mobile devices. In this study, we (1) analyze the knowledge leakage risk (KLR) caused by the use of mobile devices in knowledge-intensive Australian organizations, (2) present a conceptual research model to explain the determinants that influence KLR through the use of mobile devices grounded in the literature, (3) conduct interviews with security and knowledge managers to understand what strategies they use to mitigate KLR caused by the use of mobile devices and (4) use content analysis and the conceptual model to frame the preliminary findings from the interviews.*

# Introduction

Information and knowledge leakage has become a significant security risk to Australian organizations. A recent global research study conducted by the Ponemon Institute on leakage found each security incident in Australian business cost an average US$2.8 million and that Australian organisations spend the second most worldwide (US$1.2 million each on average) on investigating and assessing information breaches (Ponemon Institute 2016).

Recent literature shows how organizations struggle with leakage of sensitive organizational information across various avenues, such as social media, cloud computing and portable data devices (Ahmad et al. 2014, 2015; Jiang et al. 2013; Krishnamurthy and Wills 2010; Mohamed et al. 2006). Although much of the literature has focused on technical aspects of leakage (i.e., data and information), scant research has been conducted on knowledge leakage through mobile devices in particular (Agudelo et al. 2015; Ghosh and Rai 2013; Zahadat et al. 2015).

Although the use of mobile devices (whether employee or organization owned) has shown to be convenient, this convenience comes at a high security cost. The use of such devices by knowledge workers in knowledge-sharing activities poses a problem for confidentiality. Challenges in confidentiality occur as a result of employee's security (mis)behaviours. Therefore, the focus should shift from technological (e.g., firewall, antivirus, and compartmentalization) and formal (i.e., policies, standards and procedures) controls to human factors (Agudelo et al. 2015, 2016; Ahmad et al. 2014).

Whether deliberately, or inadvertently, workers are usually more responsible than hackers for breaches of information. In fact, research has shown that the culture and people within an organisation are just as likely to be the source of data leakage (Ahmad et al. 2015; Colwill 2009; Crossler et al. 2013; Nuha, Molok, and Ahmad 2011). For example, confidential and sensitive information is sometimes shared inadvertently through social media and mobile devices (Agudelo et al. 2016; Nuha, Molok, Ahmad, et al. 2011).

In this empirical study, we address the gap found in the literature and in practice by undertaking an exploratory study conducting a series of interviews with information security and knowledge experts in a number of knowledge-intensive organizations in Australia, addressing the following research question:

- *How can knowledge-intensive organizations in Australia mitigate the knowledge leakage risk (KLR) caused by the use of mobile devices?*
    - *What strategies are used by knowledge-intensive organizations in Australia?*

As this paper is part of a bigger study, we will address the specific sub-question: *What strategies are used by knowledge-intensive organizations in Australia?*

To answer this question, this paper takes a contextual approach to understand how risk changes depending on the circumstances within which knowledge leakage occurs and uses a research conceptual model to explain the factors that influence the risk of knowledge leakage through the use of mobile devices. Understanding the determinants behind this risk can assist organizations in developing more effective formal (policy), informal (culture, behavior, Security Education Training and Awareness) and technological controls that can address this issue (Dhillon 2007).

The rest of this paper is structured as follows: First, it provides salient concepts found in the key background literature. Second, the conceptual model is shown followed by a brief discussion of the constructs, third, the initial analysis of the interviews follows. Finally, the study outlines potential contributions and future work.

# Key Literature

## *Knowledge*

There is abundant literature addressing the difference between data, information and knowledge in sources such as Boisot and Canals (2004) and Dahlbom and Mathiassen (1993). Boisot and Canals (2004) state that raw data becomes information when individuals are able to add meaning from such data, and, adding the contextual understanding in conjunction with the background of such data allows knowledge to be inferred. Therefore, knowledge is intertwined with data and information. Consequently, the leakage of knowledge is also related to the leakage of data and information. This distinction is important for our study because from the leakage of data/information, leakage of knowledge may occur just by drawing on inference, that is, we gain knowledge by inference – the process of inferring things based on what is already known (Dahlbom and Mathiassen 1993).

This study adopts the definition of knowledge given by Davenport and Prusak (1998):

*"Knowledge is a fluid mix of framed experience, values, contextual information, and expert insight that provides a framework for evaluating and incorporating new experiences and information. It originates and is applied in the minds of knowers. In organizations, it often becomes embedded not only in documents or repositories but also in organizational routines, processes, practices, and norms."*

According to this definition, knowledge is complex, a mixture of various elements, intuitive and therefore, hard to capture. Moreover, knowledge is embedded in people, and as such, may be unpredictable and intangible. Knowledge derives from information and to turn information into knowledge, human mediation is required (Davenport and Prusak 1998). Although knowledge is further divided into tacit (present in employee's minds) and explicit (knowledge that has been codified into artefacts) (Nonaka 1991), from the perspective of mobile devices, this study will only focus on explicit knowledge leakage, since its disclosure is more likely to occur in mobile device settings than tacit knowledge leakage, such as when key personnel leave the organization to a competitor (Frishammar et al. 2015).

Information and knowledge have become key strategic assets (Bollinger and Smith 2001) for knowledge-intensive organizations to achieve sustained competitive advantage, innovation and value creation (Nonaka and Toyama 2003; Sveiby 1997). Similarly, MacDougall and Hurst (2005) contend that the adoption of knowledge workers, employees who produce value by utilizing their knowledge rather than physical labor, allows organizations to develop their knowledge assets. These individuals perform work based on their information assets for the coordination and management of organizational activities (Sorensen et al. 2008). Ristovska et al. (2012) also focus on the importance of knowledge embedded in knowledge workers as it is an organizational asset for achieving sustainable competitive advantage which can be materialized into documentation and organizational processes. The importance of expertise in organizations relies heavily on exercising specialist knowledge and competencies, or alternatively, the management of organizational competencies and capabilities which belong to employees or knowledge workers (Blackler 1995; Thompson and Walsham 2004).

Knowledge, in this sense is the information residing in the mind of the knowledge worker, personalized by the individual based on facts, procedures, concepts, interpretations, ideas, observations and judgments which is codified into artefacts such as documentation, processes and guidelines (Alavi and Leidner 2001).

## *Knowledge Leakage*

Knowledge leakage (KL) is defined in this paper: as the **accidental** or **deliberate** loss or unauthorized transfer of organizational knowledge intended to stay within a firm's boundary resulting in the deterioration of competitiveness and industrial position of the organization (Frishammar et al. 2015; Nunes et al. 2006).

According to the knowledge leakage definition, KL can occur from the disclosure of sensitive details, information or data as meaning can be inferred by a competitor based on understanding of context and leveraged even further to generate insights and advance their own competitiveness to the detriment of the organization's competitive advantage (Ahmad et al. 2015; Annansingh 2012; Davenport and Prusak 1998; Molok et al. 2010).

Although knowledge loss due to a lack of knowledge management procedures is also defined as knowledge leakage (Nunes et al. 2006), the focus in this study will be on knowledge leakage directly or indirectly caused by knowledge workers when performing knowledge work through mobile devices, particularly, the accidental loss derived from misbehaviours (failing to comply policy and procedures), as it is considered the most challenging channel of leakage for organizations to control (Nunes et al. 2006). The inadvertent loss, caused by insiders can be influenced by addressing human behaviour habits through policy, culture and awareness as opposed to malicious insiders who are deliberately seeking to leak knowledge/information (Colwill 2009) and are not influenced by such controls. Therefore, the focus on this study will be on addressing unintentional leakage caused by non-malicious insiders.

Drawing upon the standard definition of risk, knowledge leakage risk (KLR) is defined as the probability that KL occurs multiplied by the impact of the KL to the organization (Ahmad et al. 2015), i.e.,

$$KLR = Probability\ of\ KL \times Impact\ of\ KL\ to\ the\ Organization$$

## *Contexts*

In order to address the issue of knowledge leakage risk through mobile devices in organizations, this paper takes a context-specific perspective to understand how risk changes according to the circumstances and factors within which leakage occurs.

Although knowledge leakage is enabled by the employee in control of the mobile device, there are multiple environmental factors that affect the use of mobile devices for knowledge work. Nonaka and Toyama (2003) suggest that knowledge creation, sharing and distribution are achieved through the interactions between the individual, the organization and the environment. The environment influences the individual while, at the same time, individuals are continuously recreating their environment through their social interactions. This proposes that social factors in human interactions constantly change the environment in which knowledge is created. Nonaka and Toyama (2003) developed a model of knowledge creation in order to explain the conversion of knowledge through interactions between individuals, groups of individuals, organizations and the environment. This model not only highlights the importance of the environmental and organizational circumstances around an individual, it also highlights the importance of the social environment where individuals interact within groups to obtain information (Nonaka and Konno 1998; Nonaka and Toyama 2003).

These environments are referred to in the literature, from a mobile device perspective, as the "context" of the mobile device usage (Abdoul Aziz Diallo 2012; Chen and Nath 2008; Schilit et al. 1994). Table 1 summarises the mobile-usage-context taxonomy found in the literature. Understanding the different contexts of mobile device usage in these different settings (technological, environmental, organizational, social and personal) is important to assess the overall security risk of the device as the potential enabler of or medium through which knowledge leakage can occur in conjunction with the user and the environment within which the device is used. The importance of mobile device contexts stems from the fact that without the context within which knowledge leakage occurs, it is not possible to determine the level of risk (Benítez-Guerrero et al. 2012; Bradley and Dunlop 2005).

Mohamed et al. (2006) found that one of the key routes of knowledge leakage is *people* through social contexts of mobile usage. These routes include training courses, collaborations with universities, multi-disciplinary teams and temporary workers. Through social interactions in these different contexts, knowledge is shared or accessible to other users. Social context also includes the use of social networking platforms on mobile devices (Krishnamurthy and Wills 2010).

Due to the nature of mobile device usage, the context of a device usage transitions across many changes in technical, social and locational environments (technological, environmental, organizational and personal contexts). Through the interactions of these dynamic contexts with one another, the risk of knowledge leakage also becomes dynamic. Thus, knowledge can be leaked through the technological, organizational, personal, and network context amongst others (Diallo et al. 2011, 2014). As an illustration of this phenomenon, Astani et al (2013) found that a significant amount of employees from information sensitive industries such as banking, connected their mobile devices to unsecured public Wi-Fi networks (i.e., technological context, environmental context) which exposes the device to the security vulnerabilities of those networks and may be used as a vehicle for knowledge leakage. By simply changing the network connection to a public Wi-Fi network, these employees are drastically changing the technological and environmental contexts and, therefore, their "mobile device usage context" in which the device is operating, changing the risk profile of their device, drastically affecting the potential for knowledge leakage.

| Context | Reference | Description |
|---|---|---|
| Environmental | Kofod-Petersen & Cassens (2006); Nieto, Botía, & Gómez-Skarmeta, (2006) | The environmental context is defined as the conjunction of the following contexts: temporal context, spatial context, social context, technological context, and business context |
| Personal | Kofod-Petersen & Cassens (2006) | The personal context provides the attributes of cognitive skills and draws on psychological and physiological contexts: psychological, goal, cognition, physiological, identity, actions |
| Social | Nieto et al., (2006) | Provides a social perspective of context, which captures the attributes of people (e.g. attitude, skills, and values) and the relationship of these people among each other and within the organization and collective structures. |
| Spatial | Kofod-Petersen & Cassens (2006) | Provides attributes of location and answers the question of where the interaction is conducted. The following are some constructs of this category: spatial objects, localization, location, season, weather, geography, routes, building |
| Temporal | Abdoul Aziz Diallo (2012) | Temporal context is defined in terms of when the activity is performed: absolute date (year, month, day, hour, minutes, seconds), relative date (yesterday, tomorrow, next month, next year, etc.) |
| User | Abdoul Aziz Diallo (2012) | User context extends on personal context adding the technological dimension and the mobile device from HCI (Human-Computer interaction) perspective |
| Location | Abdoul Aziz Diallo (2012) | Location context is part of the spatial context and it is defined by: places, GPS location |
| Business | Abdoul Aziz Diallo (2012) | The business context supports the decision making process by assisting in the decision maker's situation awareness cognitive process, and taking in to consideration the following aspects: indicators, objectives, partners, competitors, market |
| Technological | Abdoul Aziz Diallo (2012) | Provides the technological and technical attributes such as: network connections, infrastructure, equipment, devices and systems. It is an aggregate context which consists of other technical constituents such as spatial, user and location context. |
| Organizational | (Crossler et al. 2013; Furnell and Rajendran 2012; | Defines the social interactions within the workplace and security behaviour determined by Information Security Policies, Security Education Training and Awareness, Culture, Standards, organizational processes and |

|  | Whitman, Michael and Mattord 2011) | procedures |
|---|---|---|
| Device | (Diallo et al. 2014; Kofod-Petersen and Cassens 2006; Nieto et al. 2006) | Technological features such as device identifier and device type (i.e., laptop, tablet, smartphone) |

**Table 1. Taxonomy of Mobile Usage Contexts derived from the literature**

These contexts are relevant to the usage of the mobile device. If a user changes devices (device context, technological context), for example, then his/her overall context (user context) will change. The new device may not have the same functions as the previous one, resulting in a new number of contexts affecting the device. Since the old device is no longer used by the user, various contexts (e.g., social, user, and location contexts) no longer apply to it. This highlights the dynamic changes in knowledge leakage risks as the circumstances of how the knowledge worker uses their mobile device change.

Additionally, people and objects are constantly moving in and out of different context risks and the relevancy of these objects and people to the context are dynamic and hence the security threat of knowledge leaking is constantly changing. For example, if John is sitting in a coffee store reading his corporate emails from his tablet before heading into work (environmental, personal and technological context) and a new customer sits down behind John (social context), John's risk context has changed as the customer may potentially read John's tablet screen (shoulder surfing). John then receives a phone call (personal and social context), which introduces a new person (caller) into the context, with whom he then discusses the agenda of the morning meeting (organizational context). This change in context risk now involves the surrounding people within earshot drastically increasing the potential for knowledge leakage.

From the literature there have been many approaches to modelling the contextual information surrounding mobiles across many disciplines of Information Technology. Most of the research into the *contextual information* and *context* of mobile devices has been focused on the technical and computing issues (Benítez-Guerrero et al. 2012; Bradley and Dunlop 2005; Diallo et al. 2013; Hofer et al. 2003; Kofod-Petersen and Cassens 2006; Schilit et al. 1994).

Similarly, Hofer et al. (2003) also extended and modelled these dimensions of context into device context (e.g. device identifier and device type) and network context (e.g. network connection types) which were included as the technical context, in a more recent study, by Abdoul Aziz Diallo (2012). However, these studies failed to address the social context, neglecting the human perspective from the mobile contextual model, namely, user behaviour.

On the other hand, Chen & Nath (2008) asserted that the social context is not independent of the technical context; it is the "interaction and compatibility" between the two that determine the effectiveness of a work system. This interdependency of the social and technical context is further reflected by Bradley and Dunlop's, (2005) "Model of Context in Computer Science" which aims to illustrate the key components and characteristics of context which are present during user-computer interaction. The key idea derived from Chen and Nath's model of context is that there are multiple contexts that contribute to the mobile usage context of mobile devices.

Expanding on Chen & Nath's (2008) social context interaction framework and Bradley and Dunlop's (2005) model of context, we address the gap in the literature by modelling such contexts from the human perspective and defining a high-level construct, knowledge leakage risk through mobile devices, as a formative construct (i.e., comprised of mobile usage contexts) which in turn informs risk mitigation strategies in organizations. This conceptual model is further explained in the following section.

## The Research Model

Figure 1 depicts our proposed research conceptual model. We develop our research model by identifying key constructs based on the two models mentioned in the previous section: Chen and Nath's (2008) "social context interaction framework" and Bradley and Dunlop's (2005) "model of context in computer science".

The criteria to select contexts for the conceptual model were based on the social context interaction framework (Chen and Nath 2008):  1) Personal context; 2)Social context are grouped together under Human factors which refer to motivations and cognitive processes, as well as social norms that are explicit and implicit from human behaviours and social interaction; 3) Environmental context; 4) Organizational context which constitute the Enterprise factors and refer to the organizational culture and behaviour, operating environment (regulations) not only within the workplace but also outside (macro environment). Finally, the Technical factors are composed of 5) Device context; 6) Technological context and refer to the technology and information systems that enable and facilitate the adoption of technology and technical artefacts to perform knowledge-sharing activities.

The constructs have been clustered in three groups: human, organizational and technical factors as defined in the "Integrative Model of IT Business Value" based on the resource-based view of the firm (Melville et al. 2004) as this model provides a framework to understand how internal (organizational resources) and external (trading partners, competitive and macro environment) factors impact organizational performance, which in our case, relates to how the mobile device usage contexts through the formative construct (i.e., KLR) contributes to improvement of organizational information and knowledge security performance signified in the construct organizational KLR mitigation strategies. In Table 2 the propositions are listed and explained with references to the literature.

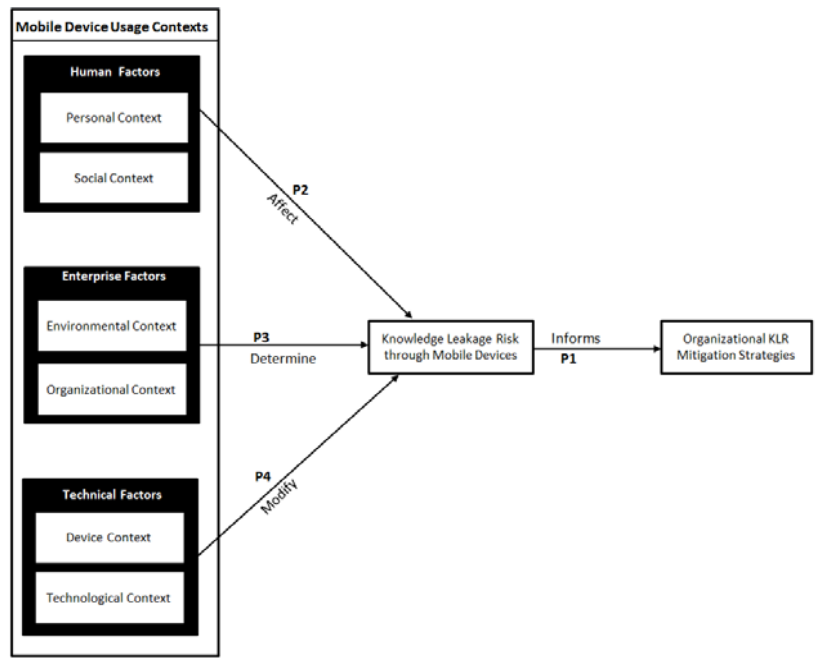

**Figure 1. Proposed Research Conceptual Model. Adapted from Bradley & Dunlop (2005), Chen & Nath (2008) and Melville et al. (2004)**

| Construct | Definition | Reference |
|---|---|---|
| Organizational KLR Mitigation Strategies | Formal, informal and technical risk control strategies, used by organizations to safeguard knowledge assets at risk. Such strategies aim to reduce risk impact or probability (risk reduction), as well as share, avoid, transfer or accept any residual risk remaining after the risk treatment. | (ISO/IEC 27005:2011 2011; Dhillon 2007) |
| Knowledge leakage Risk through Mobile Devices | Knowledge leakage risk caused by the use of mobile devices in organizations. This high-level construct will be operationalized used a qualitative scale, i.e., low, medium and high. | (27005:2011 2011; Agudelo et al. 2015; Ahmad et al. 2015) |
| Human Factors | The combination of personal and social contexts referring to individual's self-efficacy, personality traits, competences, behaviour, attitude, cognitive capabilities, motivations, experiences (personal context) as well as group's competences, social norms, peer's influence and superior's influence (social context). | (Ajzen 1991; Bandura 1978; Bradley and Dunlop 2005; Chen and Nath 2008; Melville et al. 2004) |
| Enterprise Factors | The combination of environmental and organizational contexts referring to external conditions (e.g., competitors, industry, external locations) as well as internal organizational resources and capabilities (e.g., policies, culture, processes, routines). | (Ajzen 1991; Bradley and Dunlop 2005; Chen and Nath 2008; Melville et al. 2004) |
| Technical Factors | The combination of device and technological contexts referring to the infrastructure and technological resources internal and external to the organization that enable and support knowledge-sharing activities. | (Ajzen 1991; Bradley and Dunlop 2005; Chen and Nath 2008; Melville et al. 2004) |

**Table 2. Definition of constructs in the research conceptual model**

| Proposition | Description | Explanation |
|---|---|---|

| P1 | Knowledge leakage risk through mobile devices informs the organizational knowledge leakage risk mitigation strategies. | Organizational risk mitigation strategies are designed according to the risk exposure and risk appetite (risk profile) of the organization. For instance, military organizations will have a different risk profile and, therefore, different mitigation strategy as compared to a not-for-profit organization (Baskerville et al. 2014) |
|---|---|---|
| P2 | Human factors affect the knowledge leakage risk through mobile devices. | This proposition refers to constructs such as self-efficacy, motivations, attitudes, personality traits, social norms, peer influence that affect the behaviour and perception of users when interacting with IS (Ajzen et al. 1991) |
| P3 | Enterprise factors determine the knowledge leakage risk through mobile devices. | This refers to organizational resources (Barney 1991) such as organizational structure, policies and rules, workplace practices, and culture which condition the perception of IS phenomena (Melville et al. 2004; Sveen et al. 2009; Whitman, Michael and Mattord 2011). |
| P4 | Technological factors modify the knowledge leakage risk through mobile devices. | Technological controls such as firewall, IDS, and compartmentalization lead to a diminished KL risk, particularly, when neglecting human aspects such as intentions and compliance (Bulgurcu et al. 2010; Herath and Rao 2009; Pahnila et al. 2007). |

**Table 3. Conceptual Model Propositions**

## *Knowledge Leakage Risk & Organizational Knowledge Leakage Risk Mitigation Strategies*

The resource-based view (RBV) highlights the importance of protecting resources and capabilities to sustain competitive advantage in organizations (Leonard-barton 1992). In this perspective, the organizational knowledge capability needs to be protected, and the knowledge leakage risk (KLR) associated to this asset requires assessment. However, the risk evaluation process is a subjective exercise that leads to a perceived KLR characterized by the impact and likelihood of leakage happening (ISO/IEC 27005:2011 2011). As a result, the risk treatment involves selecting one or more options for modifying the risk impact or probability. Such treatment includes implementing controls and strategies to address the residual risks that are suited to the risk profile of the organization, environment and resources. Such arguments lead to our first proposition, which serves as the foundation to examine the role of perceived KLR to firm's strategy in developing organizational mitigation controls:

**Proposition 1**. *The knowledge leakage risk through mobile devices informs the organizational knowledge leakage risk mitigation strategies.*

This proposition aims to answer our research question: *How does the knowledge leakage risk through mobile devices inform organizations' mitigation strategies?*

Human, Enterprise and Technological Factors

As discussed previously and expanding on RBV and contextual framework, previous studies have evaluated a considerable number of organizational characteristics as determinants of competitive advantage, which in turn have been classified within the broader category of basic competences or influencing factors (Chen and Nath 2008; Leonard-barton 1992):

1. Human factors, which include among other things, a firm's knowledge and skills, accumulated either through training of its workforce (Teece 2007) or as a result of the experience acquired over time (Ristovska et al. 2012). Individual competences (personal context) as well as group competences (social context) are part of the key internal capacities of a firm to develop capabilities. However, such competencies also affect the decision making and critical protection action processes (Ajzen 1991) such as risk evaluation. The above discussion leads to:

**Proposition 2.** *Human factors (**personal** and **social** contexts) affect the knowledge leakage risk through mobile devices*

2. Enterprise factors, which include the organizational resources (internal conditions) such as structure, policies, rules, workplace practices and culture (organizational context); and the environment (external conditions) in which the organization operates and interacts with other organizations including its market, regulations, competitors and external resources (environmental context) (Melville et al. 2004). These factors determine the risk profile and appetite of the organization, which leads to our third proposition:

**Proposition 3.** *Enterprise factors (**environmental** and **organizational** contexts) determine the knowledge leakage risk through mobile devices*

3. Technical factors, which include infrastructure, shared technology, system integration, technology services across and outside the organization (technological context) (Melville et al. 2004) in conjunction with mobile devices (device context) used by workers to perform their activities (Chen and Nath 2008) enable the sharing and creation of knowledge. Although, the technical capabilities facilitate the conditions for knowledge accumulation, it can also pose a challenge for knowledge protection due to the excessive reliance on technical controls to safeguard organizational knowledge assets, leading to a false sense of security. Hence we propose:

**Proposition 4.** Technological factors (**device** and **technological** contexts) modify the knowledge leakage risk through mobile devices.

## Methodology and Data Collection

Given the explorative nature of this study, we followed a qualitative research design using different participants. Data collection comprised 19 interviews to information security and knowledge managers of medium to large knowledge-intensive organizations in Australia (see tables 4 and 5). Supplementary documentation (policies and procedures) provided by the organizations was examined for triangulation. The purpose of each interview was to establish how each organization encouraged the flow and sharing of knowledge, particularly through mobile devices, while also ensuring that knowledge leakage does not occur. The reason we interviewed senior managers was to identify what strategies they had in place to prevent knowledge leakage. Interviews were conducted over a period of six months. Each interview lasted approximately 1 hour and was audio-recorded with the consent of each interviewee and transcribed verbatim and shared with each interviewee to check validity and verify the content. Additionally, documents such as policies and procedure were also analyzed to get a better understanding of knowledge protection mechanisms used in the organizations. The transcribed data was analyzed using selective, axial and thematic content analysis (Krippendorff 1980; Miles & Huberman 1994) and drawing on the different mobile contexts outlined in the research model to classify collected evidence. The Findings in terms of the different knowledge protection controls and mechanisms used by security and knowledge professionals in knowledge-intensive organizations are summarized in table 6.

| ID | Role | Industry | Experience (Years) |
|---|---|---|---|
| CIO1 | Chief Information Officer | Government | 10+ |
| SM1 | Security Manager | Banking | 15+ |
| CISO1 | Chief Information Security Officer | Consultancy | 20+ |
| SM2 | Security Manager | IT Provider | 10+ |
| CTO1 | Chief Technical Officer | IT Services | 15+ |
| SM3 | Security Manager | Insurance | 10+ |
| CISO1 | Chief Information Security Officer | Health Care | 10+ |
| CSM | Cyber Security Manager | Consultancy | 15+ |
| CISO2 | Chief Information Security Officer | Telecommunications | 15+ |
| CIKO* | Chief Information and Knowledge Officer | Government | 10+ |

**Table 4. Information Security Managers participants Information**

| ID | Role | Industry | Experience (Years) |
|---|---|---|---|
| CKO1 | Chief Knowledge Officer | Food | 15+ |
| KM1 | Knowledge Manager | Health Care | 10+ |
| KM2 | Knowledge Manager | Government | 10+ |
| CKO2 | Chief Knowledge Officer | Consultancy | 15+ |
| KM3 | Knowledge Manager | Consultancy | 10+ |
| CKO3 | Chief Knowledge Officer | Government | 15+ |
| KM4 | Knowledge Manager | Health Care | 5+ |
| KM5 | Knowledge Manager | Education | 10+ |
| KM | Knowledge Manager | Not-for-Profit | 15+ |
| CIKO* | Chief Information and Knowledge Officer | Government | 10+ |

**Table 5. Information Knowledge Managers participants Information**

*This participant was considered in both groups due to his expertise in both fields.*

### *Findings*

The findings in this section list the leakage strategies reported by security and knowledge managers of knowledge-intensive organizations in Australia that were part of this study. Given that our sample was specific and relatively small, it is not generalizable. However, it is indicative of the practices followed by many organizations in Australia.

In summary, the findings of this study show many of the strategies overlap the human, enterprise and technical factors. However, security and knowledge managers in knowledge-intensive organizations in Australia are aware of the differences and the focus is now shifting from the enterprise and technical dimension to the human factors. Organizations are increasingly aware of the current threat landscape, and the only way to keep current and maintain competitive advantage in an environment that is ever changing and increasingly complex, is through protection of resources and organizational capabilities, such as knowledge and information assets.

Similarly, strategies that focus on employees' behaviors as well as, behavioral change are just as important. The key is to increase employees' understanding and awareness that the way they interact with other people, their mobile devices and computing systems can enhance or diminish the effectiveness of a security program. Sustained periodic training and awareness also serve to reinforce policies and procedures in the minds of employees.

By analyzing the data collected from the interviews using content analysis and drawing on the research model, we have provided a framework to classify the evidence and a tool for practitioners that can be used as a guide and checklist to combat knowledge and information leakage caused by the use of mobile devices in knowledge-intensive organizations in Australia.

| Factor | Context | Type of Control | Leakage Strategy reported by |
|---|---|---|---|
| Human | Personal | Strategic | -Corporate Governance<br>-Security Risk Management Framework (Identification of Knowledge assets and processes)<br>-Legal Framework<br>-Whistle Blower Policy<br>-Security Culture Strategy<br>-Enterprise Mobility Strategy<br>-Security Indicators for training and behaviors<br>-Knowledge Communities & Portal<br>-Knowledge/Information leakage Response Plan |
| | | Operational | -On-going HR Screening Processes<br>-Background checks<br>-Personnel Risk Assessment<br>-Awareness & Training Workshops for mobile users<br>-Simulations & Games<br>-Key Performance Indicators linked to training<br>-Behavior Profiling (AI)<br>-Use of BigData and Machine learning technologies to analyze Mobile User interactions |
| | Social | Strategic | -Security Culture Strategy<br>-Knowledge Communities<br>-Peer Mentoring Policy<br>-Whistle Blower Policy |
| | | Operational | -Group risk Assessments<br>-Information/Knowledge Owners<br>-Separation of Concerns<br>-Need- to- know principle<br>-Gamification<br>-Knowledge Directory |
| Enterprise | Organizational | Strategic | -Corporate Governance<br>-Enterprise Mobility Strategy<br>-Knowledge Management Strategy<br>-Risk Management Framework<br>-Security Culture Strategy<br>-Legal Framework<br>-Knowledge/Information Leakage Response Plan<br>-Multi-disciplinary integration among areas (HR, IT, Legal) Strategy |
| | | Operational | -Knowledge and Information security procedures<br>-Information Classification system<br>-Document Tagging<br>-Non-Disclosure Agreements, Patents,<br>-Informal protection mechanisms: Misinformation, Secrecy, Fast Innovation cycle<br>-Mobile procedures in place (Loss, theft, breaches) |
| | Environment | Strategic | -Strategic Liaisons between organizations ( Government, Research, private and public sector)<br>-Market analysis (Fast Innovation Cycle)<br>-Competitor/Adversary Analysis (Tactics, Motivation, Goals)<br>-Environment Risk Policy (for mobile workers out of the office)<br>-Mobile Device use Profiling out of the office (AI – Machine Learning) |
| | | Operational | -Procedures for use of Mobile Devices in external environments<br>-Secure external communication/knowledge channels with partners<br>-Legal Framework: Partnerships/Agreements with external parties<br>-Intelligence gathering (Adversaries, competitors) |
| Technical | Device | Strategic | -Enterprise Mobility Strategy<br>-Mobile Device management<br>-Security Policy |
| | | Operational | -End point Security<br>-Encryption<br>-Compartmentalization (Personal/work space)<br>-Device use profiling (BigData, telemetry, Machine learning)<br>-Mobile Policy enforcement |
| | Technological | Strategic | -Enterprise Mobility Strategy<br>-Knowledge Management Strategy (Portal, communities, Wikis)<br>-Security Operation Center (SOC)<br>-System Incident and Event management (SIEM)<br>-Security Policy<br>-Security Response Plan |
| | | Operational | -Technical Security Policy enforcement<br>-Mobile Device Patching policy<br>-User profiling in terms of network, infrastructure and mobile device usage (BigData, Machine Learning) |

**Table 6. Summary of knowledge leakage strategies observed in the study.**

## Conclusion and Future Directions

The results of this exploratory study are expected to have both practical and theoretical implications. This study is expected to contribute to IS security research by proposing a comprehensive conceptual model which will be empirically tested in later phases and will investigate the determinants of knowledge leakage risk through mobile devices in knowledge-intensive organizations operating in highly competitive environments in Australia. Our study is also expected to provide meaningful implications for security and knowledge managers in organizations to improve risk mitigation strategies associated to knowledge leakage.

In today's security landscape, mobile devices present some new threats to organizations' security and knowledge management strategy. Effective KLR mitigation strategies will help organizations better manage those devices in their environment protecting their organizational knowledge. This study is the first attempt to view KLR through mobile devices in organizations from a mobile usage perspective using a contextual approach combining human, enterprise and technological dimensions. By analyzing the determinants that influence the knowledge leakage risk through mobile devices in organizations, addressing not only technological aspects but also human and organizational aspects, the proposed model presents a better way to design mitigation strategies and leakage risk controls (i.e., formal, informal and technological) that is more likely to be accepted and followed by employees (Dhillon 2007).

We have proposed a conceptual model which will be tested at a later stage and seeks to explain how the knowledge leakage risk is influenced by human, enterprise and technological factors and how such KLR informs organizations' mitigation strategies. Empirical confirmation and refinement of the research conceptual model is an important future research direction to follow. Additionally, we have categorized the findings of the initial interviews on leakage strategies used by knowledge-intensive organizations in Australia drawing on the conceptual proposed.

Our study has the following limitations. First, our sample was specific and relatively small. The findings in this study need to be explored in larger empirical studies across multiple organizational sectors. Second, our main source of information was interviews with senior-level managers. As such, we did not explore leakage-related behaviors at the operational level in terms of operational staff, which, in turn, points to the need for further studies in terms of actual employee behavior.

This study is the first part of a bigger study. In this current phase, we surveyed security and knowledge managers about the different strategies and mechanisms used to address knowledge leakage caused by mobile devices in different contexts. In this phase, 19 interviews with knowledge managers (9) and security experts (10) were conducted. The objective of this phase was to better conceptualize the different constructs and investigate how such factors characterize KLR mitigation strategies using the conceptual model as reference.

In a future second phase, we will conduct two focus groups, one with knowledge managers and other with security managers from different knowledge-intensive sectors in Australia to further improve the concepts and the underlying propositions in the model. The goal of this phase is to develop specific-sector insights (i.e., private, military, governmental and not-for-profit organizations) in order to contrast different industries.

In a third phase, we will conduct in-depth multiple case studies, following Yin' s (2003) methodology, in different types of knowledge-intensive organizations (e.g., private, military, governmental and not-for-profit organizations) which operate in highly-competitive environments in order to validate our findings and the proposed conceptual model from our previous phase and further refine the model. The objective of this phase is to generalize the findings of this research to other industries.

# References

27005:2011, I. 2011. *ISO 27005: 2011 Information technology--Security techniques--Information security risk management. ISO.*

Abdoul Aziz Diallo, B. 2012. "Mobile and Context-Aware GeoBI Applications: A Multilevel Model for Structuring and Sharing of Contextual Information," *Journal of Geographic Information System*, pp. 425–443 (doi: 10.4236/jgis.2012.45048).

Agudelo, C. A., Bosua, R., Ahmad, A., and Maynard, S. B. 2015. "Understanding Knowledge Leakage & BYOD ( Bring Your Own Device ): A Mobile Worker Perspective," *The 26th Australasian Conference on Information Systems - ACIS2015*, pp. 1–13.

Agudelo, C. A., Bosua, R., Ahmad, A., and Maynard, S. B. 2016. "Mitigating Knowledge Leakage Risk in Organizations through Mobile Devices: A Contextual Approach," in *27th Australasian Conference on Information Systems 2016 - ACIS2016*, pp. 1–12.

Ahmad, A., Bosua, R., and Scheepers, R. 2014. "Protecting organizational competitive advantage: A knowledge leakage perspective," *Computers and Security* (42), Elsevier Ltd, pp. 27–39 (doi: 10.1016/j.cose.2014.01.001).

Ahmad, A., Bosua, R., Scheepers, R., and Tscherning, H. 2015. "Guarding Against the Erosion of Competitive Advantage: A Knowledge Leakage Mitigation Model," *international Conference on Information Systems ICIS2015*, pp. 1–13.

Ajzen, I. 1991. "The theory of planned behavior," *Organizational Behavior and Human Decision Processes* (50:2), pp. 179–211 (doi: 10.1016/0749-5978(91)90020-T).

Ajzen, I., Netemeyer, R., and Ryn, M. Van. 1991. "The theory of planned behavior," *Orgnizational Behavior and Human Decision Processes* (50), pp. 179–211 (doi: 10.1016/0749-5978(91)90020-T).

Alavi, M., and Leidner, D. E. 2001. "Review: Knowledge Management and Knowledge Management Systems: Conceptual Foundations and Research Issues," *MIS quarterly* (25:1), pp. 107–136.

Annansingh, F. 2012. *Exploring the Risks of Knowledge Leakage : An Information Systems Case Study Approach*, INTECH Open Access Publisher.

Astani, M., Ready, K., and Tessema, M. 2013. "BYOD issues and strategies in organizations," *Issues in Information Systems* (14:2), article, , pp. 195–201.

Bandura, A. 1978. "Self-efficacy: Toward a unifying theory of behavioral change," *Advances in Behaviour Research and Therapy* (1:4), pp. 139–161 (doi: 10.1016/0146-6402(78)90002-4).

Baskerville, R., Spagnoletti, P., and Kim, J. 2014. "Incident-centered information security: Managing a strategic balance between prevention and response," *Information and Management* (51:1), Elsevier B.V., pp. 138–151 (doi: 10.1016/j.im.2013.11.004).

Benítez-Guerrero, E., Mezura-Godoy, C., and Montané-Jiménez, L. G. 2012. "Context-aware mobile collaborative systems: Conceptual modeling and case study," *Sensors* (12:10), article, Molecular Diversity Preservation International, pp. 13491–13507.

Blackler, F. 1995. "Knowledge, knowledge work and organizations: An overview and interpretation,"

*Organization studies* (16:1995), pp. 1021--1046.
Boisot, M., and Canals, A. 2004. "Data, information and knowledge: have we got it right?," *Journal of Evolutionary Economics* (14:1), pp. 43–67 (doi: 10.1007/s00191-003-0181-9).
Bollinger, A. S., and Smith, R. D. 2001. "Managing organizational knowledge as a strategic asset," *Journal of Knowledge Management* (5:1), pp. 8–18 (doi: 10.1108/13673270110384365).
Bradley, N. A., and Dunlop, M. D. 2005. "Toward a Multidisciplinary Model of Context to Support Context-Aware Computing," *Human-Computer Interaction* (20:4), pp. 403--446 (doi: 10.1207/s15327051hci2004).
Bulgurcu, B., Cavusoglu, H., and Benbasat, I. 2010. "Information security policy compliance: an empirical study of rationality-based beliefs and information security awareness," *MIS quarterly* (34:3), article, , pp. 523–548.
Chen, L., and Nath, R. 2008. "A socio-technical perspective of mobile work," *Information, Knowledge, Systems Management* (7:1), pp. 41–60.
Colwill, C. 2009. "Human factors in information security: The insider threat - Who can you trust these days?," *Information Security Technical Report* (14:4), Elsevier Ltd, pp. 186–196 (doi: 10.1016/j.istr.2010.04.004).
Crossler, R. E., Johnston, A. C., Lowry, P. B., Hu, Q., Warkentin, M., and Baskerville, R. 2013. "Future directions for behavioral information security research," *Computers & Security* (32), Elsevier Ltd, pp. 90–101 (doi: 10.1016/j.cose.2012.09.010).
Dahlbom, B., and Mathiassen, L. 1993. *Computers in context: The philosophy and practice of systems design*, Blackwell Publishers, Inc.
Davenport, T. H., and Prusak, L. 1998. "Working knowledge: how organizations manage what they know," *Harvard Business Press* (31:4), pp. 1–15 (doi: 10.1109/EMR.2003.1267012).
Dhillon, G. 2007. *Principles of Information Systems Security: text and cases*, Wiley New York, NY.
Diallo, B. A. A., Badard, T., Hubert, F., and Daniel, S. 2011. "Towards context awareness mobile Geospatial BI (GeoBI) applications," in *International Cartography Conference (ICC). Paris*, inproceedings, .
Diallo, B. A. A., Badard, T., Hubert, F., and Daniel, S. 2013. "Context-based mobile GeoBI: enhancing business analysis with contextual metrics/statistics and context-based reasoning," *GeoInformatica*, pp. 1–29 (doi: 10.1007/s10707-013-0187-x).
Diallo, B. A. A., Badard, T., Hubert, F., and Daniel, S. 2014. "Context-based mobile GeoBI: Enhancing business analysis with contextual metrics/statistics and context-based reasoning," *GeoInformatica* (18:2), pp. 405–433 (doi: 10.1007/s10707-013-0187-x).
Frishammar, J., Ericsson, K., and Patel, P. C. 2015. "The dark side of knowledge transfer: Exploring knowledge leakage in joint R&D projects," *Technovation* (42), pp. 75–88 (doi: 10.1016/j.technovation.2015.01.001).
Furnell, S., and Rajendran, A. 2012. "Understanding the influences on information security behaviour," *Computer Fraud and Security* (2012:3), pp. 12–15 (doi: 10.1016/S1361-3723(12)70053-2).
Ghosh, A., and Rai, P. K. G. S. 2013. "Bring Your Own Device (Byod): Security Risks and Mitigating Strategies," *Journal of Global Research in Computer Science* (4:4), pp. 62–70.
Herath, T., and Rao, H. R. 2009. "Protection motivation and deterrence: a framework for security policy compliance in organisations," *European Journal of Information Systems* (18:2), article, Nature Publishing Group, pp. 106–125.
Hofer, T., Schwinger, W., Pichler, M., Leonhartsberger, G., Altmann, J., and Retschitzegger, W. 2003. "Context-awareness on mobile devices - the hydrogen approach," *Proceedings of the36th Annual Hawaii International Conference on System Sciences, 2003.* (43:7236) (doi: 10.1109/HICSS.2003.1174831).
Jiang, X., Li, M., Gao, S., Bao, Y., and Jiang, F. 2013. "Managing knowledge leakage in strategic alliances: The effects of trust and formal contracts," *Industrial Marketing Management* (42:6), Elsevier Inc., pp. 983–991 (doi: 10.1016/j.indmarman.2013.03.013).
Kofod-Petersen, A., and Cassens, J. 2006. "Using activity theory to model context awareness," in *Modeling and Retrieval of Context*, incollection, Springer, pp. 1–17.
Krishnamurthy, B., and Wills, C. E. 2010. "On the leakage of personally identifiable information via online social networks," *ACM SIGCOMM Computer Communication Review* (40:1), p. 112 (doi: 10.1145/1672308.1672328).
Leonard-barton, D. 1992. "Core Capabilities and Core Rigidities - a Paradox in Managing New Product Development," *Strategic Management Journal* (13), pp. 111–125 (available at <Go to ISI>://WOS:A1992JF55700008).
MacDougall, S. L., and Hurst, D. 2005. "Identifying tangible costs, benefits and risks of an investment in intellectual capital: Contracting contingent knowledge workers," *Journal of Intellectual Capital* (6:1), pp. 53–71 (doi: 10.1108/14691930510574663).
Melville, N., Kraemer, K., and Gurbaxani, V. 2004. "Review: Information Technology and Organizational Performance: an Integrative Model of It Business Value," *MIS Quarterly* (28:2), pp. 283–322 (doi: 10.2307/25148636).
Mohamed, S., Mynors, D., Grantham, A., Walsh, K., and Chan, P. 2006. "Understanding one aspect of the knowledge leakage concept: people," in *Proceedings of the European and Mediterranean Conference on Information Systems (EMCIS)*, pp. 2–12 (doi: 10.1504/IJEB.2007.012974).
Molok, N. N. a., Ahmad, A., and Chang, S. 2010. "Understanding the factors of information leakage through online social networking to safeguard organizational information," *ACIS 2010 Proceedings - 21st Australasian Conference on Information Systems* (available at http://www.scopus.com/inward/record.url?eid=2-s2.0-84870387706&partnerID=tZOtx3y1).
Nieto, I., Botía, J. A., and Gómez-Skarmeta, A. F. 2006. "Information and hybrid architecture model of the OCP contextual information management system," *Journal of Universal Computer Science* (12:3), article, , pp.

357–366.
Nonaka, I. 1991. "The knowledge-creating company," *Harvard Business Review* (85:7–8) (doi: 10.1016/0024-6301(96)81509-3).
Nonaka, I., and Konno, N. 1998. "The concept of 'ba': Building a foundation for knowledge creation," *California management review* (40:3), pp. 40–54.
Nonaka, I., and Toyama, R. 2003. "The knowledge-creating theory revisited: knowledge creation as a synthesizing process," *Knowledge Management Research & Practice* (1:1), pp. 2–10 (doi: 10.1057/palgrave.kmrp.8500001).
Nuha, N., Molok, A., and Ahmad, A. 2011. "Disclosure of Organizational Information by Employees on Facebook: Looking at the Potential for Information Security Risks," *ACIS 2011 Proceedings* (1:1), pp. 1–12.
Nuha, N., Molok, A., Ahmad, A., and Chang, S. 2011. "Exploring the use of online social networking by employees: Looking at the potential for information leakage," *Pacific Asia Conference on Information Systems - PACIS*, p. 138.
Nunes, M. B., Annansingh, F., Eaglestone, B., and Wakefield, R. 2006. "Knowledge management issues in knowledge-intensive SMEs," *Journal of Documentation* (62:1), pp. 101–119 (doi: 10.1108/00220410610642075).
Pahnila, S., Siponen, M., and Mahmood, A. 2007. "Employees' behavior towards IS security policy compliance," in *System Sciences, 2007. HICSS 2007. 40th Annual Hawaii International Conference on*, inproceedings, , p. 156b--156b.
Ponemon Institute. 2016. "2016 cost of data breach study: global [Traverse City, Michigan, USA].,"
Ristovska, M., Popovska, Z., and Stankosky, M. 2012. "Managing Knowledge Assets for Competitiveness in the Knowledge Era," *The George Washington University* (1:1), pp. 1–6.
Schilit, B., Adams, N., and Want, R. 1994. "Context-aware computing applications," in *Mobile Computing Systems and Applications, 1994. WMCSA 1994. First Workshop on*, inproceedings, , pp. 85–90.
Sorensen, C., Al-Taitoon, A., Kietzmann, J., Pica, D., Wiredu, G., Elaluf-Calderwood, S., Boateng, K., Kakihara, M., and Gibson, D. 2008. "Exploring enterprise mobility: Lessons from the field," *Information, Knowledge, Systems Management* (7:1), pp. 243--271 (available at https://web-b-ebscohost-com.ezp.lib.unimelb.edu.au/ehost/pdfviewer/pdfviewer?sid=12919e4d-823a-4f5e-9505-ac4a5e667793%40sessionmgr111&vid=1&hid=123).
Sveen, F. O., Torres, J. M., and Sarriegi, J. M. 2009. "Blind information security strategy," *International Journal of Critical Infrastructure Protection* (2:3), Elsevier B.V., pp. 95–109 (doi: 10.1016/j.ijcip.2009.07.003).
Sveiby, K. E. 1997. *The new organizational wealth: Managing \& measuring knowledge-based assets*, Berrett-Koehler Publishers (available at http://ptarpp2.uitm.edu.my/silibus/neworgan.pdf).
Teece, D. J. 2007. "Explicating Dynamic Capabilities: The Nature and Microfoundations of (Sustainabile) Enterprise Performance," *Strategic Management Journal* (298:13), pp. 1319–1350.
Thompson, M., and Walsham, G. 2004. "Placing Knowledge Management in Context," *Journal of Management Studies* (41:July), pp. 725–747.
Whitman, Michael and Mattord, H. 2011. *Principles of information security*, Cengage Learning.
Zahadat, N., Blessner, P., Blackburn, T., and Olson, B. a. 2015. "BYOD security engineering: a framework & its analysis," *Computers & Security* (doi: 10.1016/j.cose.2015.06.011).